\def\epsfpreprint{Y}   
                       
                       \if \epsfpreprint Y
\documentstyle[12pt,epsf]{article} \fi \if \epsfpreprint N
\documentstyle[12pt]{article} \fi

\def\figure#1#2#3{\if \epsfpreprint Y \epsfxsize=#3 truein
\centerline{\epsffile{fig_#1.eps}}
\centerline{\vbox{{\bf \noindent Figure #1.} #2}}
\bigskip \fi}

\def\Psibar{\overline{\Psi}}

\def\qbar{\overline{q}}

\def\slash{\!\!\!\!/}

\def\meff{m_{\rm eff}}
\def\bfm#1{\mbox{\boldmath $#1$}}

\def\fs{\ \ \ .}

\def\betat{\bar{\beta_1}}
\def\betatc{{\bar{\beta_1}}_c}
\def\betatpc{{\bar{\beta_1}}_{c}}
\def\mfpc{{m_f}_{c}}

\def\spose#1{\hbox to 0pt{#1\hss}}
\def\ltapprox{\mathrel{\spose{\lower 3pt\hbox{$\mathchar"218$}}
 \raise 2.0pt\hbox{$\mathchar"13C$}}}
\def\gtapprox{\mathrel{\spose{\lower 3pt\hbox{$\mathchar"218$}}
 \raise 2.0pt\hbox{$\mathchar"13E$}}}
\def\inapprox{\mathrel{\spose{\lower 3pt\hbox{$\mathchar"218$}}
 \raise 2.0pt\hbox{$\mathchar"232$}}}

\def\figsizea{6.4}
\def\figsizeb{4.0}

\def\one{ 
The large N critical value $\betatc$ versus $m_0$ 
for a $6^3$ lattice with antiperiodic boundary conditions along the
time direction.
}
\def\two{ 
The real part of the maximum and minimum large N decay rates as a
function of $m_0$ for a $6^3$ lattice with antiperiodic boundary
conditions along the time direction.
}
\def\three{
The large N $\sigma_s$ as a function of $L_s$ for a $6^3$ lattice with
antiperiodic boundary conditions along the time direction, $m_f=0$ and
various $m_0$ at $\betat = 0.3$ above the transition.  The diamonds,
squares, crosses, circles, and stars correspond to $m_0 = 0.2, 0.4,
0.6, 0.8$, $1.0$.
}
\def\four{
The large N phase boundary of the parity-flavor broken phase of the 
$SU(2) \times SU(2)$ model on a $6^3$ lattice with antiperiodic 
boundary conditions along the time direction with $m_0=1$ and for
various $L_s$ values. From bottom to top the $L_s$ values are $2$,
$3$, $4$, and $5$. The parity-flavor symmetry is broken
inside the oval-looking regions.
}
\def\five{
For fixed $\betat=0.05$ the origin $-\mfpc$ (diamonds) and width
$\delta \mfpc$ (circles) of the large N parity-flavor broken phase of
the $SU(2) \times SU(2)$ model is plotted versus $L_s$ for $m_0=1$ and
for a $6^3$ lattice with antiperiodic boundary conditions along the
time direction.
}
\def\six{$<\sigma>$ as a function of $\betat$ for the 
$Z_2 \times Z_2$ model on
a $6^3$ lattice with antiperiodic boundary conditions along the time
direction, $m_f=0.02$, $L_s=12$, $m_0=0.4$ (diamonds), $m_0=1.0$
(squares) and $m_0=1.6$ (crosses).  The solid lines are the large N
predictions for the same parameters and the dotted lines are the large
N predictions for $m_f=0$ and $L_s=\infty$.
}
\def\seven{
The decay and decay rate of $<\sigma>$ as a function of $L_s$ for the
$Z_2 \times Z_2$ model on a $6^3$ lattice with antiperiodic boundary
conditions along the time direction and $m_f=0.02$.  In figure 7a
$m_0=1.0$ and $\betat=0.05$ (diamonds), $\betat=0.1$ (squares) and
$\betat=0.25$ (crosses).  In figure 7c $\betat=0.05$ and $m_0=0.6$
(diamonds), $m_0=0.8$ (squares), $m_0=1.0$ (crosses), and $m_0=1.1$
(circles).  The solid lines in figures 7a and 7c are fits to $c_0 +
c_1 e^{-c_2 L_s}$ and the dotted lines are the large N
predictions. The decay rates $c_2$ from the fits in figures 7a and 7c
are shown in figures 7b and 7d respectively.  The solid lines in these
figures are the large N predicted minimum and maximum decay rates.
}
\def\eight{
The average value of the third component of the pion field versus
$m_f$ for the $SU(2) \times SU(2)$ model on
a $6^3$ lattice with antiperiodic boundary conditions along
the time direction, $\bfm{h} = (0,0,0.1)$, $\betat=0.05$, $L_s=2$ and
$m_0=1.0$.  The ``outer'' solid line is the large N prediction for
$\bfm{h} = (0,0,0.1)$ and the ``inner'' one for $\bfm{h} =
(0,0,0)$. The diamonds are the results of the numerical simulations.
}
\begin{document}

\title{\bf Fermion--scalar interactions with domain wall fermions} 
\vskip 1. truein
\author{P. Vranas, I. Tziligakis and J. Kogut
\\
\\
Physics Department \\
University of Illinois \\
Urbana, IL 61801
\\
\\
}
\maketitle

\begin{abstract}

Domain wall fermions are defined on a lattice with an extra direction
the size of which controls the chiral properties of the theory.  When
gauge fields are coupled to domain wall fermions the extra direction
is treated as an internal flavor space. Here it is found that this is
not the case for scalar fields. Instead, the interaction takes place
only along the link that connects the boundaries of the extra
direction. This reveals a richness in the way different spin particles
are coupled to domain wall fermions.  As an application, 4-Fermi
models are studied using large N techniques and the results are
supported by numerical simulations with N=2. It is found that the
chiral properties of domain wall fermions in these models are good
across a large range of couplings and that a phase with parity-flavor
broken symmetry can develop for negative bare masses if the number of
sites along the extra direction is finite.

\end{abstract}

\newpage

\section{Introduction}
\label{sec_intro}

Domain wall fermions \cite{Kaplan} provide an alternative to standard
lattice fermions and have already been used to formulate lattice gauge
theories (for reviews see \cite{DWF_reviews} and references
therein). Here domain wall fermions are used to study lattice theories
of fermions and scalars. Such theories can be studied analytically
using large N methods and applications of these theories requiring
enhanced control over the chiral symmetries may be of interest. For
example, such applications may involve 4-Fermi models or models with
Higgs fields interacting via Yukawa couplings.

Domain wall fermions (DWF) are defined on the sites of a space-time
lattice with one extra direction. The dimension of this lattice will
be denoted as $d+1$. In the method of \cite{Furman_Shamir} the
boundary conditions along the extra direction are free and as a result
light fermion surface states develop on the boundary.  The plus
chirality components of the Dirac spinor are exponentially localized
on one wall while the minus ones are on the other. If the extra direction
has $L_s$ sites then the two chiralities overlap only by an amount that
is exponentially small in $L_s$. Therefore, the extra parameter $L_s$
controls the amount by which the regulator breaks the chiral symmetry
at any lattice spacing. As a result, and contrary to staggered or
Wilson fermions, the chiral and continuum limits are
separated. Furthermore, in numerical applications the cost for
recovery of the chiral symmetry is only linear in $L_s$.  These
properties make DWF an attractive regulator in problems where good
control of the chiral symmetries is needed.  For example, DWF have been
used in studies of the finite temperature QCD phase transition
\cite{lat98_PMV,lat98_Fleming,lat98_Kaehler,ICHEP_NHC,dpf99_PMV}.

One would have expected that the formulation of theories with DWF
coupled to scalar fields should closely follow that of DWF coupled to
gauge fields. However, this is not the case and as a result an
interesting difference between the way the scalar and gauge fields are
coupled to DWF emerges. The formulation and a discussion of this
difference is presented in section \ref{sec_interaction}. Two 4-fermi
models, one with a $Z_2 \times Z_2$ and one with an $SU(2) \times
SU(2)$ chiral symmetry in three dimensions, are studied in section
\ref{sec_lrg_n} using large N techniques. In section \ref{sec_hmc}
these models are studied using Hybrid Monte Carlo simulations with
$N=2$. Conclusions are presented in section
\ref{sec_conclusions}. 

Using DWF in 4-Fermi models was also suggested in
\cite{Bitar_private}. For a recent implementation of overlap fermions
to the gauged Gross-Neveu model see \cite{Ichinose}.

\section{The fermion--scalar interaction}
\label{sec_interaction}

The free DWF action in the formulation of \cite{Furman_Shamir} is:
\begin{equation}
S = - \sum_{x,x^\prime,s,s^\prime} \Psibar(x,s) D(x,s; x^\prime,
s^\prime) \Psi(x^\prime,s^\prime)
\label{action}
\end{equation}
with Dirac operator
\begin{equation}
D(x,s; x^\prime,s^\prime) = \delta(s-s^\prime) D\slash(x, x^\prime)
+ D\slash^\bot(s,s^\prime) \delta(x-x^\prime)
+ E^\bot(s,s^\prime;x) \delta(x-x^\prime)
\label{Dirac_op}
\end{equation}
where
\begin{equation}
D\slash(x, x^\prime) = 
{1\over 2} \sum_{\mu=1}^{d} \left[ (1+\gamma_\mu)
\delta(x+\hat\mu - x^\prime) + (1-\gamma_\mu)
\delta(x^\prime+\hat\mu - x) \right]  \nonumber \\
+ (m_0 - d)\delta(x-x^\prime) 
\label{Dslash_prl_free}
\end{equation}

\begin{equation}
D\slash^\bot(s,s^\prime) = 
\left\{ \begin{array}{ll} 
P_R \delta(1-s^\prime) - \delta(s - s^\prime) & s=0 \\ 
P_R \delta(s+1 - s^\prime) + P_L \delta(s-1 - s^\prime) - \delta(s-s^\prime) & 0 < s < L_s-1 \\ 
P_L \delta(L_s-2 - s^\prime) - \delta(s - s^\prime) & s = L_s -1
\end{array}
\right.
\label{Dslash_perp_free}
\end{equation}

\begin{equation}
E^\bot(s,s^\prime) = 
- \ m_f \ [ Q_R(s,s^\prime) + Q_L(s,s^\prime)]
\label{e_perp_free}
\end{equation}

\begin{equation}
P_{R,L} = { 1 \pm \gamma_5 \over 2}
\label{prl}
\end{equation}

\begin{equation}
Q_R(s,s^\prime) = P_R \delta(L_s-1 - s)  \delta(0-s^\prime), \ \ \  
Q_L(s,s^\prime) = P_L \delta(0-s) \delta(L_s-1 - s^\prime) 
\label{qrl}
\end{equation}

In the above equations $m_0$ is a five-dimensional mass representing
the ``height'' of the domain wall. 
Among other things, the parameter $m_0$ controls the number of light
species \cite{Kaplan}. For example, in four dimensions the theory is symmetric
around $m_0=5$ and for $0< m_0 < 2$ corresponds to a theory with one
light species, for $2< m_0 < 4$ to a theory with four and for 
$4< m_0 < 6$ to a theory with six \cite{lat98_RDM}.
The parameter $m_f$ explicitly mixes the two
chiralities and as a result it controls the bare fermion mass of the
$d$-dimensional effective theory in a linear fashion. The DWF Dirac operator satisfies
the identity \cite{Furman_Shamir}:
\begin{equation}
\gamma_5 R D_F \gamma_5 R = D_F^\dagger
\label{reflection}
\end{equation}
with $R$ the reflection operator along the fifth direction. As a
result, the two species determinant is real and non-negative
\begin{equation}
\det{D_F}^2 = \det{D_F^\dagger D_F}
\label{dwf_det}
\end{equation}
From a calculation of the propagator or from the lowest eigenvalue
of the fermion matrix the fermion mass for the one flavor
free theory is \cite{PMV}:
\begin{equation}
\meff = m_0 (2 - m_0) \left[ m_f + (1-m_0)^{L_s}\right], \ \ \ \ 0 < m_0 < 2 
\label{meff_free}
\end{equation}

From the free action one can see that there is more than one way one
could couple a gauge field to DWF in a gauge invariant way.  The most
straightforward choice of introducing a gauge field is to also define
it in $d+1$ dimensions and couple it in the standard gauge invariant
way. However, it turns out that this choice does not lead to the
correct theory in $d$ dimensions.  The correct way to couple the gauge
field was arrived at by thinking of the extra dimension as yet another
internal flavor space \cite{Kaplan,NN1}.  Then one has to introduce the
gauge field only in the $d$ dimensional space and couple it the same
way to all fermion ``flavors''. Obviously, this also results to a
manifestly gauge invariant formulation. 
In particular, the Dirac operator is as in
eq. \ref{Dirac_op} but with $D\slash$ defined as:
\begin{eqnarray}
D\slash(x,x^\prime) &=& 
{1\over 2} \sum_{\mu=1}^{d} \left[ (1+\gamma_\mu)
U_\mu(x) \delta(x+\hat\mu - x^\prime) + (1-\gamma_\mu)
U^\dagger_\mu(x^\prime) \delta(x^\prime+\hat\mu - x) \right]  \nonumber \\
&+& (m_0 - d)\delta(x-x^\prime)
\label{Dslash_prl_gauge}
\end{eqnarray}

Therefore, although one can think of the extra direction as a new
space-time dimension it is more natural to think of it as an internal
flavor space.  From this point of view, if one wishes to formulate a
theory that in $d$-dimensions would be a theory of a scalar field
coupled to a fermion bilinear, one would couple the scalar field to all
internal fermion flavors the same way, as was done for the case of a
gauge field. However, this choice will not result to the desired
theory in $d$ dimensions. To see this consider the simple example of a
target theory of fermions coupled to a real valued scalar field $\sigma$
in $d$ dimensions with continuum action:
\begin{equation}
S = \int \qbar \gamma_\mu \partial_\mu q + m \qbar q 
+ \sigma \qbar q + \beta \sigma^2
\label{action_sigma_cont}
\end{equation}
If $m = 0$ this model has a $Z_2 \times Z_2$ chiral symmetry.
It is clear that a large $N$ or mean field analysis of this model 
will indicate that the propagator mass would be
$m + <\sigma>$ where $<\sigma>$ is the vacuum expectation value
of the $\sigma$ field. Clearly a lattice formulation of this
model using DWF should at least reproduce this simple result.
However, if the $\sigma$ field is coupled to all fermion degrees
of freedom the same way, as is the mass $m_0$, i.e. 
$\Psibar(x,s) \sigma(x) \Psi(x,s)$, a similar analysis
will result to a propagator mass as in eq. \ref{meff_free},
but with $m_0$ replaced by $m_0 + <\sigma>$.
This is obviously the wrong result. Even if one was to accept
this exponential behavior by redefining the $\sigma$ field
this theory would have the peculiar property of changing the number
of light species depending on the value of $\sigma$.
Clearly, in order to get the correct result one has to couple
the $\sigma$ field in the same fashion as the $m_f$ mass
and not as the $d+1$ mass $m_0$. Then the Dirac operator is as in
eq. \ref{Dirac_op} but with $E^\bot$:
\begin{equation}
E^\bot(s,s^\prime;x) = 
- [m_f + \sigma(x)] \ [Q_R(s,s^\prime) + Q_L(s,s^\prime)]
\label{e_perp_sigma}
\end{equation}

In the resulting action the $\sigma$ field couples with the fermion
fields only across the boundary links of the $d+1$ dimension.  That
this is the correct action can be seen by observing that the
interaction term in the action can be rewritten as:
\begin{equation}
\Psibar(x,L_s-1) \sigma(x) P_R \Psi(x,0) + 
\Psibar(x, 0) \sigma(x) P_L  \Psi(x, L_s-1) = 
\qbar(x) \sigma(x)  q(x)
\label{sigma_interaction}
\end{equation}
with:
\begin{eqnarray} 
q(x)    &=& P_R \Psi(x,0) + P_L \Psi(x, L_s-1) \nonumber \\
\qbar(x) &=& \Psibar(x,L_s-1) P_R + \Psibar(x, 0) P_L
\label{effective_fields}
\end{eqnarray} 
The fields $\qbar$, $q$ correspond to the light $d$-dimensional fields
\cite{Furman_Shamir}. Since these are the only propagating fields in $d$
dimensions and since, in terms of these fields the correct global symmetry
is in place, the above action should describe the desired target theory.

Another way to see that this is the correct action is to consider
adding such an interaction to Neuberger fermions
\cite{Neuberger_fermions}.  In that formulation the fermion fields are
standard $d$ dimensional fields and the interaction can be added
unambiguously. Then if the discrete finite flavor version of that
formulation is used one can directly recover the DWF action. Since the
$\sigma$ field is added as a mass term, one can see by simple
inspection that the above action is recovered. Furthermore, the
connection can also be made exact by using the results of
\cite{Kikukawa} where the $d$-dimensional action arrived at by
integrating all heavy fermion degrees of freedom is expressed solely
in terms of the $\qbar$, $q$ fields.

It should also be noted that scalar fields were used in a different
context \cite{Kaplan,chiral_scalar} in an attempt to formulate chiral
gauge theories using DWF.  There, in order to maintain gauge
invariance, the scalar fields were defined on a ``slice'' in
the extra direction.

With the interaction defined as in eq. \ref{e_perp_sigma}
the transfer matrix of the model along the $d+1$
direction \cite{NN1,Furman_Shamir} is independent of the
$\sigma$ field and is therefore the free field transfer matrix.  The
$\sigma$ field is defined only on the boundary and therefore only
affects the boundary operator $\cal O$ of
\cite{Furman_Shamir}. In the case of a gauge theory the extra fermion
degrees of freedom are heavy but numerous as the
$L_s \rightarrow \infty$ limit is taken. For this reason it was shown
that they need to be subtracted and the way to do that for each
gauge field background is to divide out a bulk factor \cite{NN1} of the
form $f(U)^{L_s}$ where $f(U)$ is some function of the gauge field $U$
determined by the transfer matrix (in the formulation of
\cite{Furman_Shamir} $f(U) = \lambda_{\rm max} \det{B}$).  
Here the same bulk factor is present but unlike the gauge theory it
does not depend on the background field. Therefore it can be factored
out from the $\sigma$ path integral and as a result it becomes an
irrelevant factor that will cancel out with the same factor coming by
dividing by the partition function whenever an expectation value is
calculated. Therefore, unlike the gauge theory, it is unnecessary to
explicitly subtract it out.

The extension to more general interaction terms is straightforward.
The interaction is written in terms of the $\qbar$, $q$ fields and
using eq. \ref{effective_fields} is transcribed for the $\Psibar$,
$\Psi$ fields.  For example, a target theory with $SU(2) \times SU(2)$
chiral symmetry has Dirac operator as in eq. \ref{Dirac_op} but with
$E^\bot$:
\begin{equation}
E^\bot(s,s^\prime;x) = 
- [m_f + \sigma(x) 
+ i \gamma_5 \bfm{\pi}(x) \cdot \bfm{\tau}] \ 
[Q_R(s,s^\prime) + Q_L(s,s^\prime)]
\label{e_perp_sigma_pion}
\end{equation}
where $\bfm{\pi(x)}$ is the pion field and $\bfm{\tau}$ are the Pauli
matrices.

The above results, although contrary to naive expectations, are
straightforward.  However, the following observations can be made:

\noindent
1) For the case of scalar fields it is not natural to interpret the
extra DWF direction as an internal flavor space since the interaction
takes place only at the boundary of the extra direction.

\noindent
2) One could have thought that the introduction of an extra dimension
would have lent some freedom in the definition of the DWF interaction
with the scalar and gauge fields. However, this appears not to be the
case. The DWF formalism and symmetry requirements seem to have
naturally forced the interaction terms to be of this specific form.

\noindent
3) A picture with some richness seems to develop with different spin
fields coupled to domain wall fermions in different ways.

\section{Large N analysis}
\label{sec_lrg_n}

The analysis in this section closely follows the analysis in
\cite{Bitar_Vranas} of the same model using Wilson fermions.  
Also see \cite{Kogut_NJL1,Kogut_NJL2,Kogut_NJL3,Stag_NJL} 
for studies of 4-Fermi models with staggered fermions.

Consider a 4-Fermi model in the continuum with
$SU(2) \times SU(2)$ chiral symmetry and Lagrangian:
\begin{equation}
{\cal L} = \qbar(i \partial\slash  - m)q +
{G_1 \over 2}\left[(\qbar q)^2 + 
(\qbar i \gamma_5 \bfm{\tau} q)^2 \right].
\label{Lagrangian1}
\end{equation}

In the above expression all indices have been suppressed. The
fermionic field $q$ is a flavor $SU(2)$ doublet and a color $SU(N)$
$N$-column vector. The Lagrangian is diagonal in ``color'', in contrast
with the full QCD Lagrangian which is diagonal in flavor. $\bfm{\tau}
=\{\tau_1,\tau_2,\tau_3\}$ are the three isospin Pauli matrices,
$\partial\slash = \gamma^\mu \partial_\mu$, and $m$ is the bare quark
mass (if $m \neq 0$ the chiral symmetry is explicitly broken).  As is
well known, a Lagrangian density that is quadratic in the fermionic fields can
be obtained by introducing a scalar auxiliary field $\sigma$ and three
pseudoscalar auxiliary fields $\bfm{\pi}=\{\pi_1, \pi_2, \pi_3\}$
coupled to the fermion bilinear:
\begin{eqnarray}
{\cal L} & = & \qbar D q - n_f \beta_1 (\sigma^2 + \bfm{\pi}^2)
\nonumber \\ 
D & = & i \partial\slash - m - \sigma 
- i \gamma_5 \bfm{\tau \cdot \pi} \fs
\label{Lagrangian2}
\end{eqnarray}
Here $n_f$ is the number of flavors and $\beta_1 = {1\over 2 n_f G_1}$.
Transcribing this to a Euclidian lattice, using DWF with
fermion-scalar interaction as described in section
\ref{sec_interaction} and an even number of ``colors'' $N$ 
one obtains:
\begin{eqnarray}
Z &=& \int[d\Psi d\Psibar d\sigma d\bfm{\pi}] e^{-S} \nonumber \\
S &=& \sum_{i=1}^{N/2}
\left\{\Psibar^i D \Psi^i + 
\Psibar^{i+N/2} D^\dagger \Psi^{i+N/2}
\right\} + n_f \beta_1 (\sigma^2 + \bfm{\pi}^2)
\label{PartFun2}
\end{eqnarray}
where $D$ is defined in eq. \ref{Dirac_op} with $E^\bot$ as in eq.
\ref{e_perp_sigma_pion} and use of the property in eq. \ref{dwf_det}
has been made.

Following the standard large $N$ analysis, the saddle point equations
(SPE) are calculated by assuming uniform saddle fields and small
fluctuations of the form:
\begin{equation}
\sigma(x) = \sigma_s + {\delta\sigma(x) \over \sqrt{N}},\ \ \
\bfm{\pi}(x) = \bfm{\pi}_s + {\delta\bfm{\pi}(x) \over \sqrt{N}} 
\label{saddle_fields}
\end{equation} 
The saddle point equations are:
\begin{eqnarray}
&& {1\over V} {\rm Re Tr}[ D_s^{-1} (Q_R + Q_L) ] + 2 n_f \betat \sigma_s = 0 
\label{spe_1} \\
&& {1\over V} {\rm Re Tr}[ D_s^{-1} (Q_R + Q_L) i \gamma_5 \bfm{\tau} ] + 2 n_f \betat \bfm{\pi}_s = 0
\label{spe_2}
\end{eqnarray}
or in terms of the $\qbar$, $q$ fields of eq. \ref{effective_fields}:
\begin{eqnarray}
&& {1\over V} {\rm Re}<\qbar q>_s + 2 n_f \betat \sigma_s = 0 
\label{spe_qq_1} \\
&& {1\over V} {\rm Re}<\qbar i \gamma_5 \bfm{\tau} q> + 2 n_f \betat \bfm{\pi_s} = 0
\label{spe_qq_2}
\end{eqnarray}
where
\begin{equation}
\betat = {\beta_1 \over N} = {1\over 2 n_f N G_1}
\label{beta_bar}
\end{equation}

The $L_s=\infty$ and the finite $L_s$ cases need to be studied separately since
each case has a different symmetry group.

\medskip
\noindent
{\bf I.} The $L_s= \infty$ case.

If $m_f=0$ then the DWF Dirac operator has exact chiral symmetry.  If
there is spontaneous symmetry breaking one is therefore free to choose
the direction of breaking at will. The standard choice is $\sigma_s
\neq 0$ and $\bfm{\pi}_s = 0$. For this choice the second SPE
eq. \ref{spe_2} is trivially satisfied since the non-trivial
flavor part of $D^{-1}_s$ is proportional to $\bfm{\tau \pi_s}$.  
This case is identical to the case of a
simpler model without $\bfm{\pi}$ fields and $n_f=1$. Such a theory has
$Z_2 \times Z_2$ chiral symmetry. Therefore, the following analysis 
can be trivially extended for the $Z_2 \times Z_2$ model.

Using the free propagator results of \cite{PMV} after some algebra
the first SPE eq. \ref{spe_1} for $L_s = \infty$ results in:
\begin{equation}
\sigma_s \betat 
- {2 (\sigma_s + m_f) \over V} 
\sum_p { z(p)^2 \over {\bar p}^2  + (\sigma_s + m_f)^2 z(p)^2} 
= 0
\label{spe_inf_ls_1}
\end{equation}
where
\begin{equation}
{\bar p}^2 = \sum_{\mu=1}^d {\sin(p_\mu)}^2
\label{dwf_pbar}
\end{equation}
\begin{equation}
z(p) = 1 - b e^{-a}
\label{z_fac}
\end{equation}
\begin{equation}
b = 1 - m_0 + \sum_{\mu=1}^d (1 - \cos(p_\mu))
\label{dwf_b}
\end{equation}
\begin{equation}
\cosh a = { 1 + b^2 + {\bar p}^2 \over 2 b}, \ \ \ \ 0 \leq a
\label{dwf_cosha}
\end{equation}

The factor $z(p)$ plays the role of ``selecting'' Brillouin zones. For
a given range of $m_0$ the factor $z(p)$ is nonzero at the orgins of
only certain Brillouin zones.  To see this observe that at the orgin
of any Brillouin zone $\bar p \approx 0$ and therefore
eq. \ref{dwf_cosha} has solution $\exp(-a) = b$ if $|b|<1$ and
$\exp(-a) = 1/b$ if $|b|>1$. Therefore, from eq. \ref{z_fac}, one can
see that $z(p) = 1 - b^2$ if $|b|<1$ and $z(p)=0$ if $|b|>1$.  For
example, if $0< m_0 < 2$ then $z(p)$ is non zero only in the zone with
momentum components around zero.  The condition for nonzero $z(p)$ at
the orgins of a Brillouin zone:
\begin{equation}
|b| < 1 \Leftrightarrow m_0 - 2 < \sum_{\mu=1}^d (1 - \cos(p_\mu)) < m_0
\label{z_cond}
\end{equation}
is the same as the condition for the existence of normalizable states \cite{Kaplan}.
For zero momentum $z({\bfm p}=0) = m_0 (2 - m_0)$.

One can see that eq. \ref{spe_inf_ls_1} for $m_f=0$ can have two solutions:
one with $\sigma_s = 0$ corresponding to a chirally symmetric phase,
and one with $\sigma_s \neq 0$ corresponding to
a phase with spontaneously broken chiral symmetry. The critical value of $\betat$ is
obtained from eq. \ref{spe_inf_ls_1} in the limit $\sigma_s \rightarrow 0$
and is given by:
\begin{equation}
\betatc = {2 \over V} \sum_p
{ z(p)^2 \over {\bar p}^2 }
\label{spe_beta_c}
\end{equation}

It is interesting to notice the rather strong dependence of $\betatc$
on $m_0$. Similarly strong dependence of the critical coupling on
$m_0$ was found for the QCD finite temperature phase transition
\cite{lat98_PMV}. As an example, $\betatc$ is plotted versus $m_0$ for
a $6^3$ lattice with antiperiodic boundary conditions along the time
direction in figure 1. In the next section such lattices will be used
to compare the large N expressions with numerical simulations. A
similar graph can be obtained for a four-dimensional lattice but it
extends from $m_0=0$ to $m_0=10$ and is symmetric around $m_0 = 5$.

\noindent
{\bf II.} The finite $L_s$ case.

When $L_s$ is finite the DWF Dirac operator breaks chiral symmetry
explicitly even for $m_f=0$.Therefore $\sigma_s \neq 0$ and the
remaining symmetry is the $SU(2)$ flavor symmetry. This symmetry can
break spontaneously resulting to a nonzero $\bfm{\pi}_s$. If this
happens then from eq. \ref{spe_qq_2} one can see that 
$<\qbar i \gamma_5 \bfm{\tau} q> \neq 0$ and the parity as well 
as flavor symmetries are broken.  This is the Aoki phase
\cite{Aoki}. This phase has also been observed for this model with
Wilson fermions \cite{Aoki_NJL,Bitar_Vranas_lat94}.
The existence of this phase for DWF was also suggested
in \cite{Narayanan_private} and 
it may be present in QCD with DWF \cite{Columbia_private}.

First consider the phase with $\bfm{\pi}_s = 0$. Again the second SPE
eq. \ref{spe_2} is trivialy satisfied and the following analysis is
also valid for the $Z_2 \times Z_2$ model at finite $L_s$.  The first
SPE eq. \ref{spe_2} is:
\begin{eqnarray}
\sigma_s \betat - {2 \over V}  \sum_p \!\!\!\!\!\! &\{& \!\!\!\! 
(m_f + \sigma_s) \left( A_0 + A_2 \right) + b A_m \nonumber \\
&+& e^{-a(L_s-1)} \left[(m_f + \sigma_s) A_m 
+ b \left( B + A_1 + A_2 \right) \right] \nonumber \\
&+& e^{-2a(L_s-1)} \left[ (m_f + \sigma_s) A_1  
+  b A_m \right] \ \ \} = 0
\label{fntls}
\end{eqnarray}
where $A_0$, $A_1$, $A_2$, $A_m$ are functions of the momentum,
$m_f+\sigma_s$ and $e^{- a L_s}$ and are the same as in \cite{PMV}
but with $m_f$ replaced by $m_f + \sigma_s$.  This equation can be
used to calculate $\sigma_s$ as a function of the other parameters of
the theory.

It is interesting to see how chiral symmetry is restored as $L_s$
increases.  As can be seen from eq. \ref{dwf_cosha} the decay
coefficient $a$ is independent of $\betat$ and it only depends on
$m_0$ and the momenta.  In figure 2 the minimum and the maximum value
of $Re(a)$ (if $b < 0$ $a$ has imaginary part $\pm i \pi$) obtained for
different momenta is plotted as a function of $m_0$ in the range $0<
m_0 < 2$. The ``spikes'' of the maximum value are the singularities
that occur when $b=0$. As can be deduced from eq. \ref{dwf_cosha} the
peak of the minimum decay occurs at $m_0 = 2 - \sqrt{2} = 0.586$ and
is $-\ln(2 - \sqrt{2}) = 0.535$.  The decay rates between the minimum
and the maximum values have no gaps.  As $L_s$ increases eventually
the only chiral symmetry violations that remain will be controlled by
the minimum decay rate unless the observable is dominated solely by
terms with decay rate close to the maximum.  Since the minimum decay
rate is approximately constant across the full range of $m_0$ there is
no valuable option of tuning $m_0$ in order to achieve better
characteristics except perhaps around $m_0 = 0.586$. As an example,
$\sigma_s$ versus $L_s$ is plotted in figure 3 for $m_f=0$ and various
$m_0$ at $\betat = 0.3$ which is above the transition for all the
$m_0$ values.  The slopes obtained from the larger $L_s$ points are
slightly faster than the minimum decay rates of figure 2 and the
largest slope is for $m_0 = 0.6$.

Next consider the phase with $\bfm{\pi}_s \neq 0$.  The first SPE is
as in the $\bfm \pi = 0$ case eq. \ref{fntls}.
However, since the non-trivial flavor part of
$D^{-1}_s$ is proportional to $\bfm{\tau \pi_s}$ one can now eliminate
$\bfm \pi_s$ from the second SPE and obtain a second non trivial equation.
These two equations can be used to determine $\sigma_s$ and $\bfm{\pi_s}$.
The full form of these equations is complicated and not particularly
illuminating. However, if only the leading order in $m_f$ and $e^{-a L_s}$
is kept these equations can be written in the form:
\begin{equation}
\betat = {2 \over V} \sum_p \left[ z^2 \over D \right] + O(1)
\label{spe_beta_c_pb}
\end{equation}
\begin{equation}
m_f = - { \sum_p \left[ e^{-a L_s} \left( {\bar p}^2 + z^2 \over D \right) \right]
\over
\sum_p \left[ z^2 \over D \right] } + O(2)
\label{spe_mf_c_pb}
\end{equation}
\begin{equation}
D = {\bar p}^2 {\left(1 + [m_f + \sigma_s] e^{-a L_s}\right)}^2
+ z^2 {\left([m_f + \sigma_s] + e^{-a L_s}\right)}^2
+ |\bfm{\pi_s}|^2 {\left(z^2 + {\bar p}^2 e^{-2 a L_s}\right)}
\label{spe_den}
\end{equation}

If $L_s$ is an even number or if $m_0$ is such that $0 \leq b(p)$ (and
therefore $0 \leq e^{-a(p)}$) for all momenta then $e^{-a L_s}$ is
positive and $m_f$ must be negative in order to have a phase with
$\bfm{\pi_s} \neq 0$. The phase boundary $\mfpc(\betat)$ can be
obtained from the above equations by setting $\bfm{\pi_s} \rightarrow
0$ and eliminating $x = m_f + \sigma_s$. From these equations it can
be deduced that $|\mfpc|$ decreases exponentially with $L_s$.
Furthermore, for a given $\betatpc$ the width of the $\bfm{\pi_s} \neq
0$ region also decreases exponentially with $L_s$.  In figure 4 the
phase boundary of the $\bfm{\pi_s} \neq 0$ phase is given for a $6^3$
lattice with antiperiodic boundary conditions along the time
direction, $m_0=1$ and for various $L_s$ values.  From bottom to top
the $L_s$ values are $2$, $3$, $4$, and $5$.  This has been calculated
using the full form of the SPE and not just the truncated form
eq. \ref{spe_beta_c_pb} and \ref{spe_mf_c_pb}. However, the two are
nearly the same for $2 < L_s$.  Also different values of $m_0$ produce
similar results.  As can be seen from figure 4 when $L_s$ is increased
the phase boundary moves to smaller $|m_f|$ with decreasing width. In
figure 5 for fixed $\betat = 0.05$ the origin of the phase boundary
$-\mfpc$ and width $\delta \mfpc$ are plotted versus $L_s$ for
$m_0=1.0$ and a $6^3$ lattice with antiperiodic boundary conditions
along the time direction. 

The effective fermion mass $m_q$ is identified from the zero 
of eq. \ref{spe_den} for momenta ${\bf p} \approx 0$ and is:
\begin{equation}
m_q = z({\bf p}=0) [ m_f + \sigma_s + (1 - m_0)^{L_s} ]
\label{fermion_mass}
\end{equation}
The continuum limit is reached at $m_q = 0$ and the lattice spacing is
set to $\alpha \sim m_q$. This point corresponds to the largest
$\betatpc$ value of the phase boundary curves in figure 4 ($m_q$ is
positive on the upper part of the phase boundary and negative on the
lower one).  The width of the Aoki phase close to $m_q=0$ can be
obtained from eq. \ref{spe_mf_c_pb}. To lowest order in $m_q$ is:
$\delta m_f = m_q {\partial m_f \over
\partial m_q}|_{m_q=0}$.  The derivative is not zero at $m_q=0$ and therefore
$\delta m_f \sim m_q \sim \alpha$.  For an analysis concerning the
width of the Aoki phase in QCD with Wilson fermions see
\cite{Sharpe}. The above features are not particular to three
dimensions and similar results have been obtained for four-dimensional
lattices.
%

Finally, it is interesting to observe that if $m_0$ is such that $b(p)
\leq 0$ for some momenta and $L_s$ is odd then the $\bfm{\pi_s} \neq
0$ phase can occur even for positive $m_f$.  If the $\bfm{\pi_s} \neq
0$ phase needs to be avoided for any $m_0$ then one should set $0 \leq
m_f$ and $L_s$ to an even number (on the other hand, if $m_0 \leq 1$
then $0 \leq b(p)$ for any momenta and any value of $L_s$ can be used).

\section{Hybrid Monte Carlo simulations}
\label{sec_hmc}

In this section standard Hybrid Monte Carlo (HMC) simulations are
performed for $N=2$. These simulations support the large $N$ results
of the previous section. 
For all simulations the trajectory length is set to $\tau=1$ and the
step size to $\delta \tau = 0.1$. The acceptance rate is $\approx 90\%$ and
the conjugate gradient inverter iterations are $\approx 50 - 100$.
Typically $100-200$ thermalization sweeps 
were followed by $300- 400$ sweeps with measurements.
The lattice size for all simulations is $6^3$ with antiperiodic
boundary conditions along the time direction. All simulations
were done on workstations.

The $Z_2 \times Z_2$ model with action as in eq. \ref{action} and
interaction term as in eq. \ref{e_perp_sigma} was simulated first.  The
initial configuration for all simulations was a uniform configuration
with $\sigma = 1$. In figure 6, $<\sigma>$ is plotted as a function of
$\betat$ for $m_f=0.02$, $L_s=12$, $m_0=0.4$ (diamonds), $m_0=1.0$
(squares) and $m_0=1.6$ (crosses).  The solid lines are the large N
predictions for the same parameters and the dotted lines are the large
N predictions for $m_f=0$ and $L_s=\infty$. The agreement with the
large $N$ predictions is good away from the critical region where the
approximation of a uniform saddle is good.

The decay and decay rate of $<\sigma>$ as a function of $L_s$ for
$m_f=0.02$ is shown in figure 7. In figure 7a $m_0=1.0$ and
$\betat=0.05$ (diamonds), $\betat=0.1$ (squares) and $\betat=0.25$
(crosses).  In figure 7c $\betat=0.05$ and $m_0=0.6$ (diamonds),
$m_0=0.8$ (squares), $m_0=1.0$ (crosses), and $m_0=1.1$ (circles).
The solid lines in figures 7a and 7c are fits to 
$c_0 + c_1 e^{-c_2 L_s}$ and the dotted lines 
are the large N predictions. The decay
rates $c_2$ from the fits in figures 7a and 7c are shown in figures 7b
and 7d respectively.  The solid lines in these figures are the large N
predicted minimum and maximum decay rates. As can be seen, the
agreement with the large $N$ predictions is good and the decay rates
are fairly independent of $\betat$ and $m_0$ and close to the minimum
predicted value.  This is in contrast to gauge theories where the
dependence of the decay rates on the gauge coupling is significant
\cite{PMV,lat98_PMV}.

Finally the $SU(2) \times SU(2)$ model with action as in
eq. \ref{action} and interaction term as in
eq. \ref{e_perp_sigma_pion} was simulated in an effort to investigate
the presence of the parity-flavor broken phase.  The lattice
Lagrangian has exact $SU(2)$ flavor symmetry and as a result when
this symmetry is spontaneously broken there will be two exactly
massless Goldstone pions. In that case the Conjugate Gradient inverter
would not converge. Furthermore, in the small lattices considered here
spontaneous breaking can not really take place and the $\bfm{\pi}$
field would always average to zero for sufficiently large statistics.
In \cite{Aoki,Aoki_Gocksch,Bitar_pb}
these problems were treated by
adding a small external field $\bfm{h}$ that breaks the $SU(2)$ flavor
symmetry and therefore not only gives mass to the two pions but also
provides a ``preferred'' orientation for the $\bfm{\pi}$ field. 
Such an external field is
used here by adding a term that is exactly the same as the $\bfm{\pi}$ 
term in eq. \ref{e_perp_sigma_pion} but with $\bfm{\pi}$ replaced by $\bfm{h}$. 
The initial configuration for all simulations was a uniform
configuration with $(\sigma, \pi_1, \pi_2, \pi_3) = (1,0,0,1)$.  In
figure 8 the average value of $\pi_3$ is plotted versus $m_f$ for a
$6^3$ lattice with antiperiodic boundary conditions along the time
direction, $\bfm{h} = (0,0,0.1)$, 
$\betat=0.05$, $L_s=2$ and $m_0=1.0$. The ``outer'' solid
line is the large N prediction for $\bfm{h} = (0,0,0.1)$ and the
``inner'' one for $\bfm{h} = (0,0,0)$. The diamonds are the results of
the numerical simulations.  As can be seen they are in fairly good
agreement with the large N results supporting the presence of this
phase. A finite volume analysis together with an analysis involving
decreasing values of $\bfm{h}$ as in \cite{Bitar_pb} is still needed in order to
unequivocally establish the presence of the parity-flavor broken phase for $N=2$,
but this is not in the scope of this paper.

\section{Conclusions}
\label{sec_conclusions}

The interaction of domain wall fermions with scalar fields was
formulated. It was found that contrary to naive expectations this
interaction takes place only at the boundary of the extra direction.
This is in contrast to the interaction of domain wall fermions with
gauge fields which is the same along the extra direction. This seems
to indicate a picture with some richness where different spin fields
couple to domain wall fermions in different ways.

Large N techniques were used to study two 4-Fermi models, one with $Z_2
\times Z_2$ and one with $SU(2)
\times SU(2)$ chiral symmetry.  It was found that at the limit of
infinite extra direction the chiral symmetry breaks spontaneously in
the standard fashion.  However, if the size of the extra direction is
finite the $SU(2) \times SU(2)$ chiral symmetry is explicitly broken
by the regulator down to flavor $SU(2)$. It was found that this
remaining flavor symmetry can break spontaneously if the bare quark
mass is negative resulting to a parity-flavor broken phase of the Aoki
type. Hybrid Monte Carlo simulations were performed for those models
with $N=2$ on $6^3$ lattices with antiperiodic boundary conditions
along the extra direction. The results were found to support the
large N predictions.


\section*{Acknowledgments}

We thank G. Fleming for useful comments.
This research was supported in part by NSF under grant \# NSF-PHY96-05199.


\vfill 
\eject

%
%
\if \epsfpreprint Y \eject
\null \vskip -0.8 truein
\figure{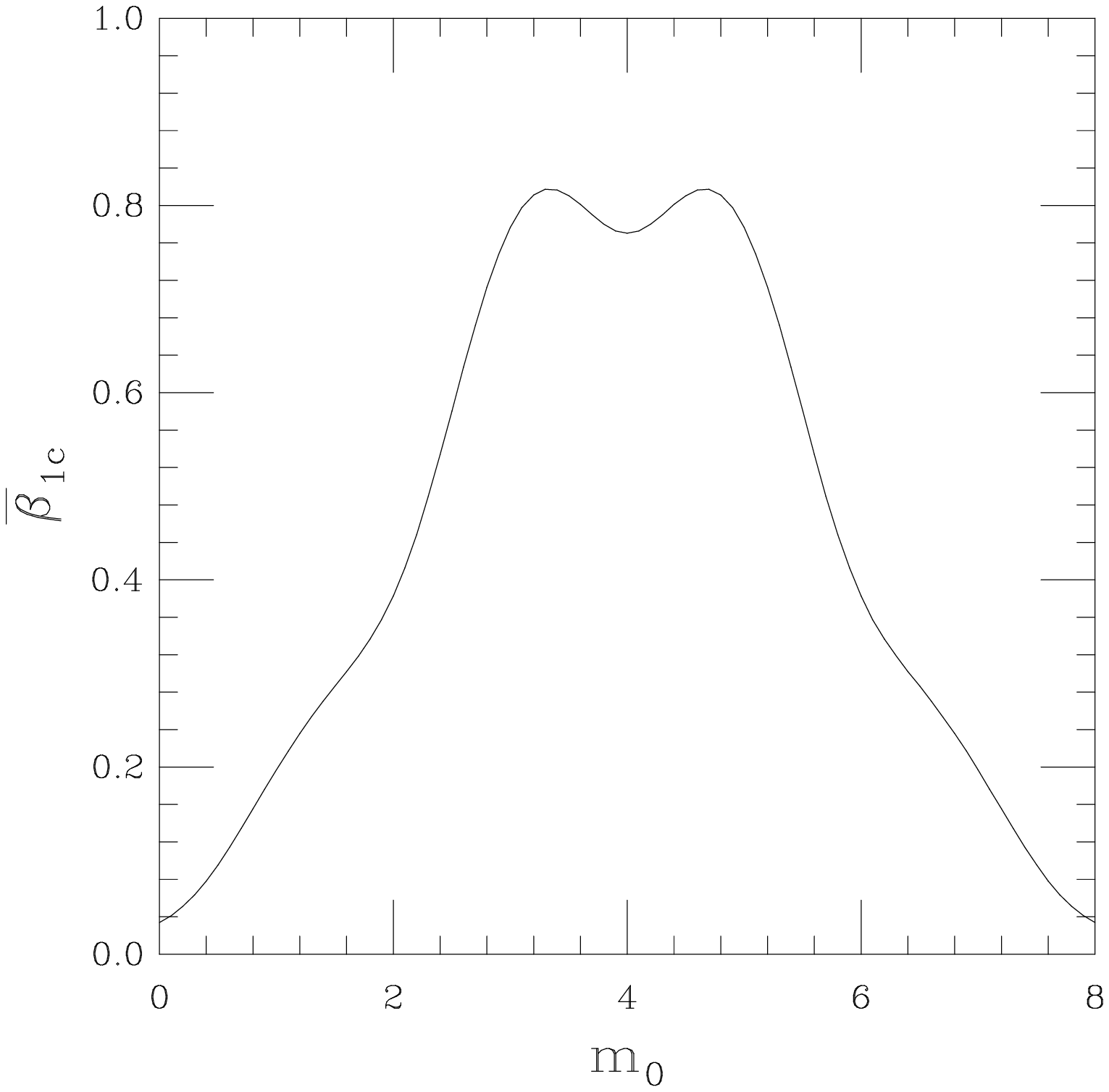}{\one}{\figsizeb} 
\null \vskip -0.8 truein
\figure{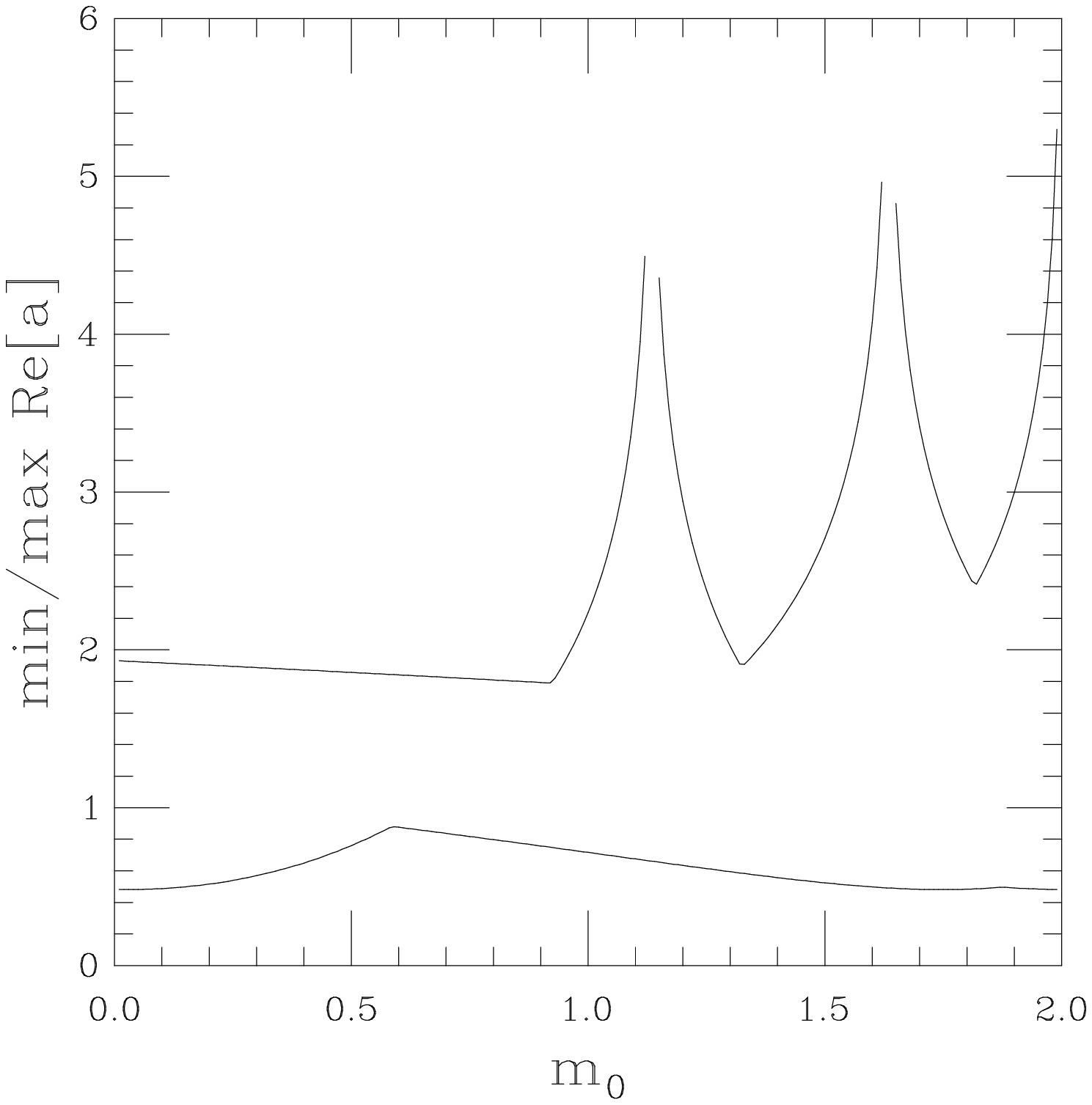}{\two}{\figsizeb} 
\null \vskip -0.8 truein
\figure{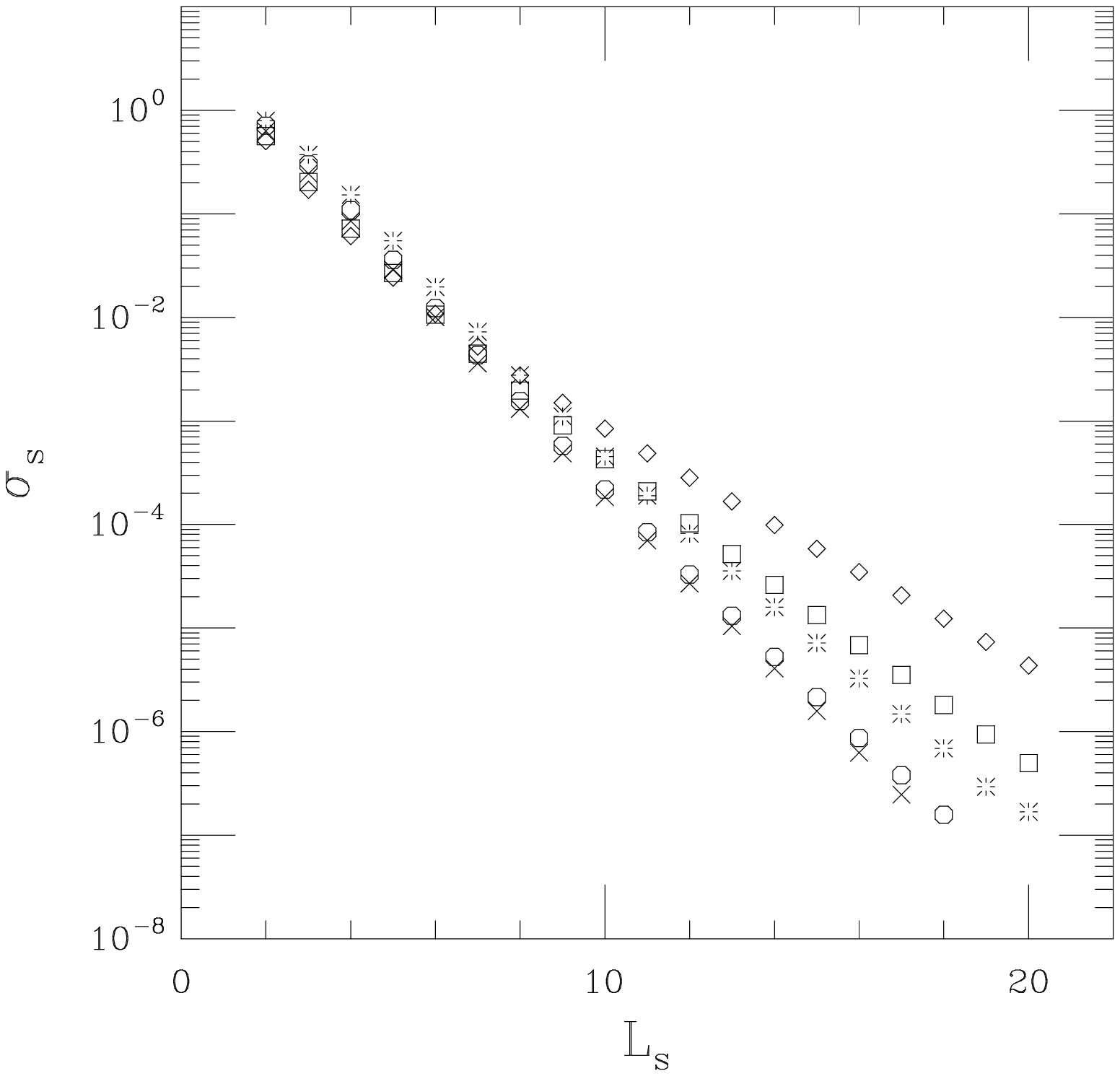}{\three}{\figsizeb} 
\null \vskip -0.8 truein
\figure{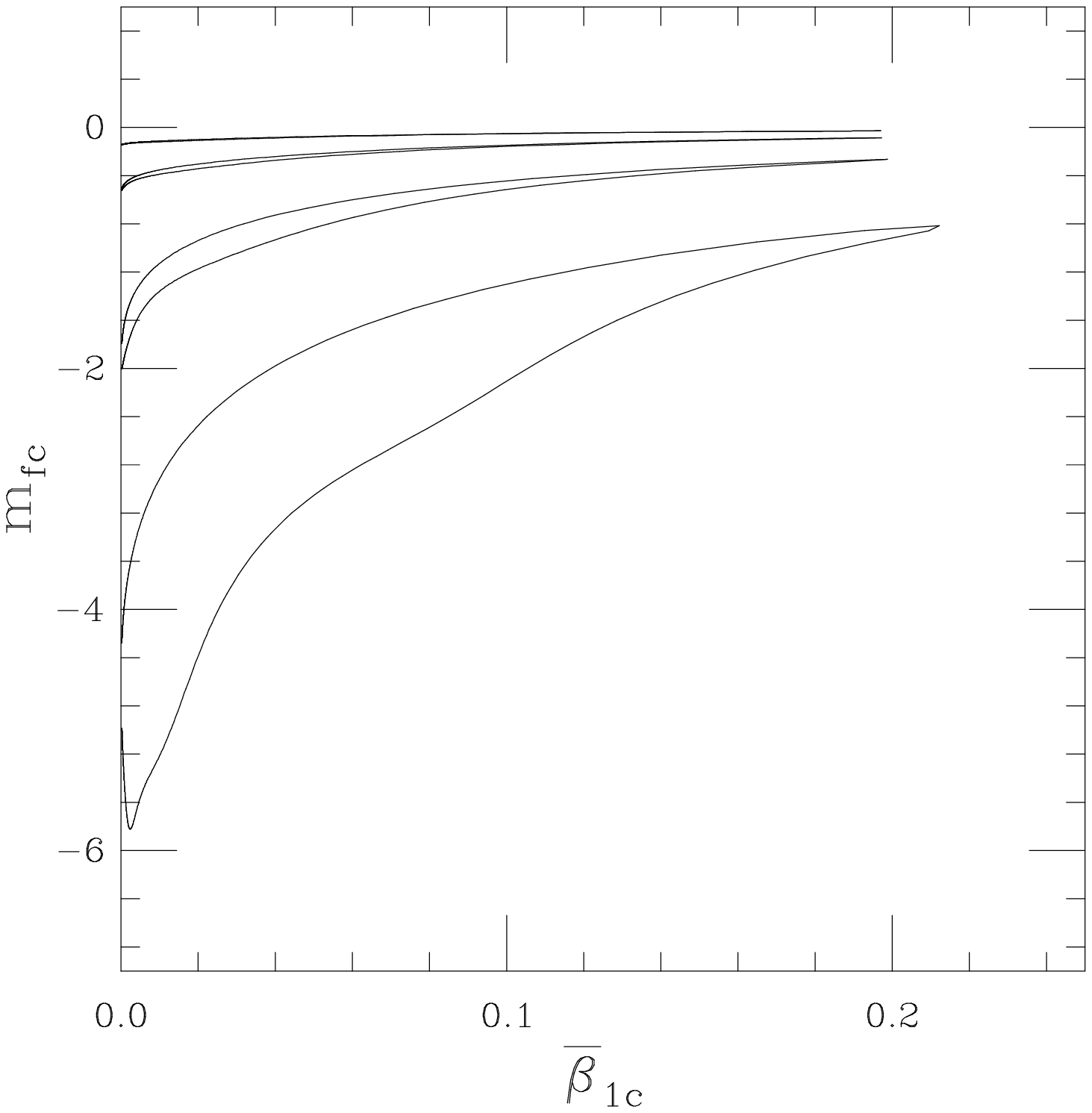}{\four}{\figsizeb} 
\null \vskip -0.8 truein
\figure{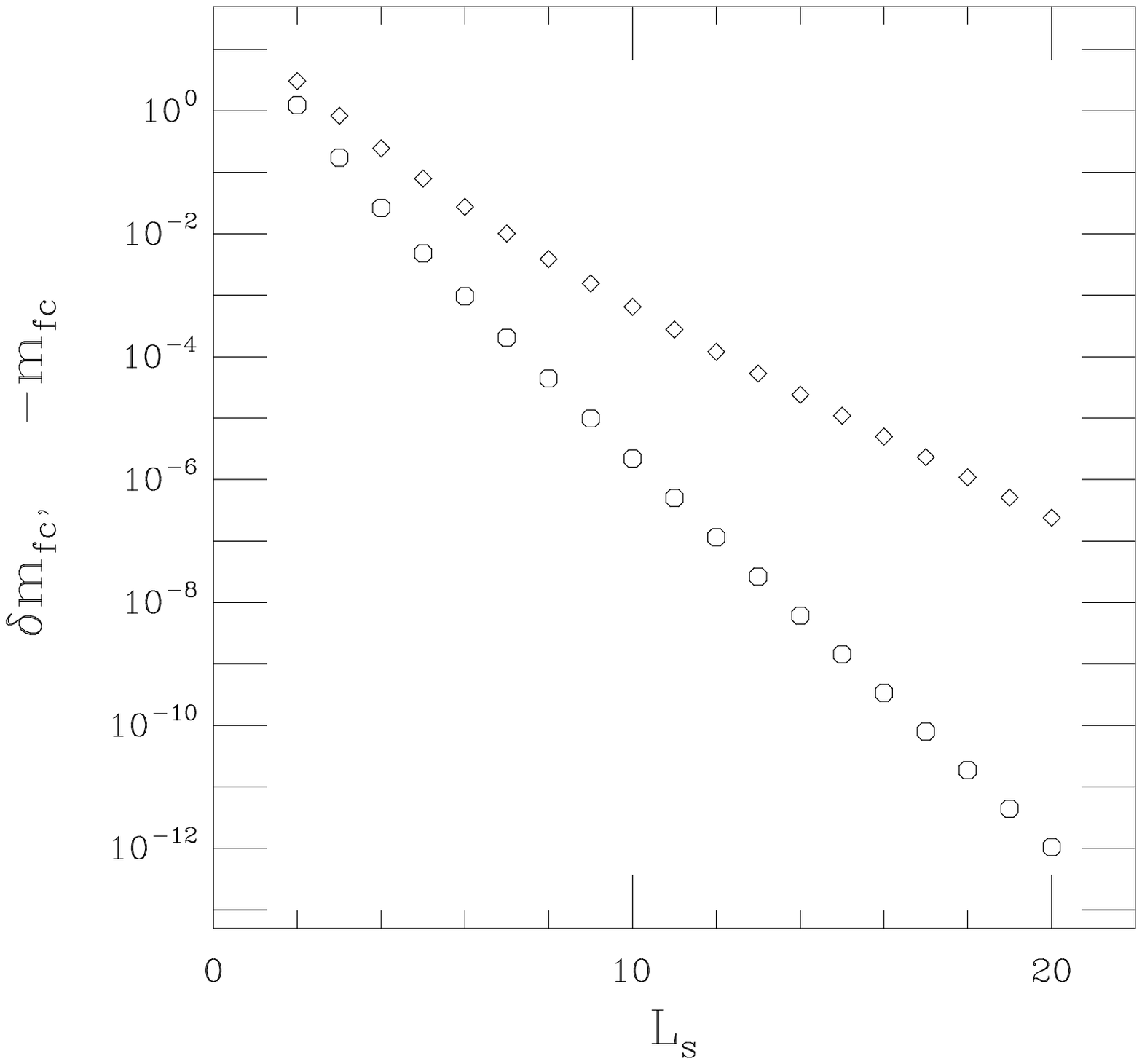}{\five}{\figsizeb} 
\null \vskip -0.8 truein
\figure{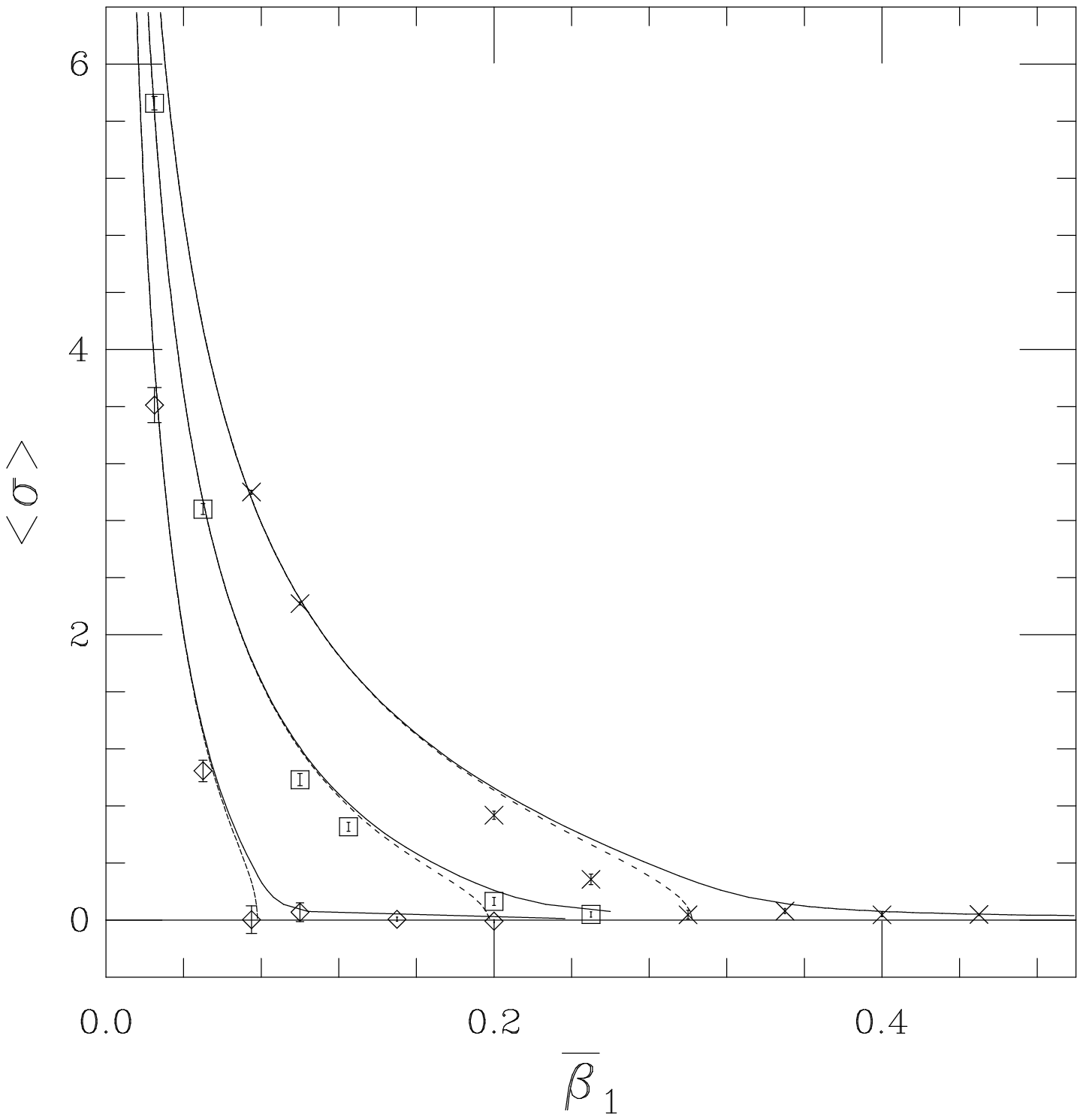}{\six}{\figsizeb} 

\figure{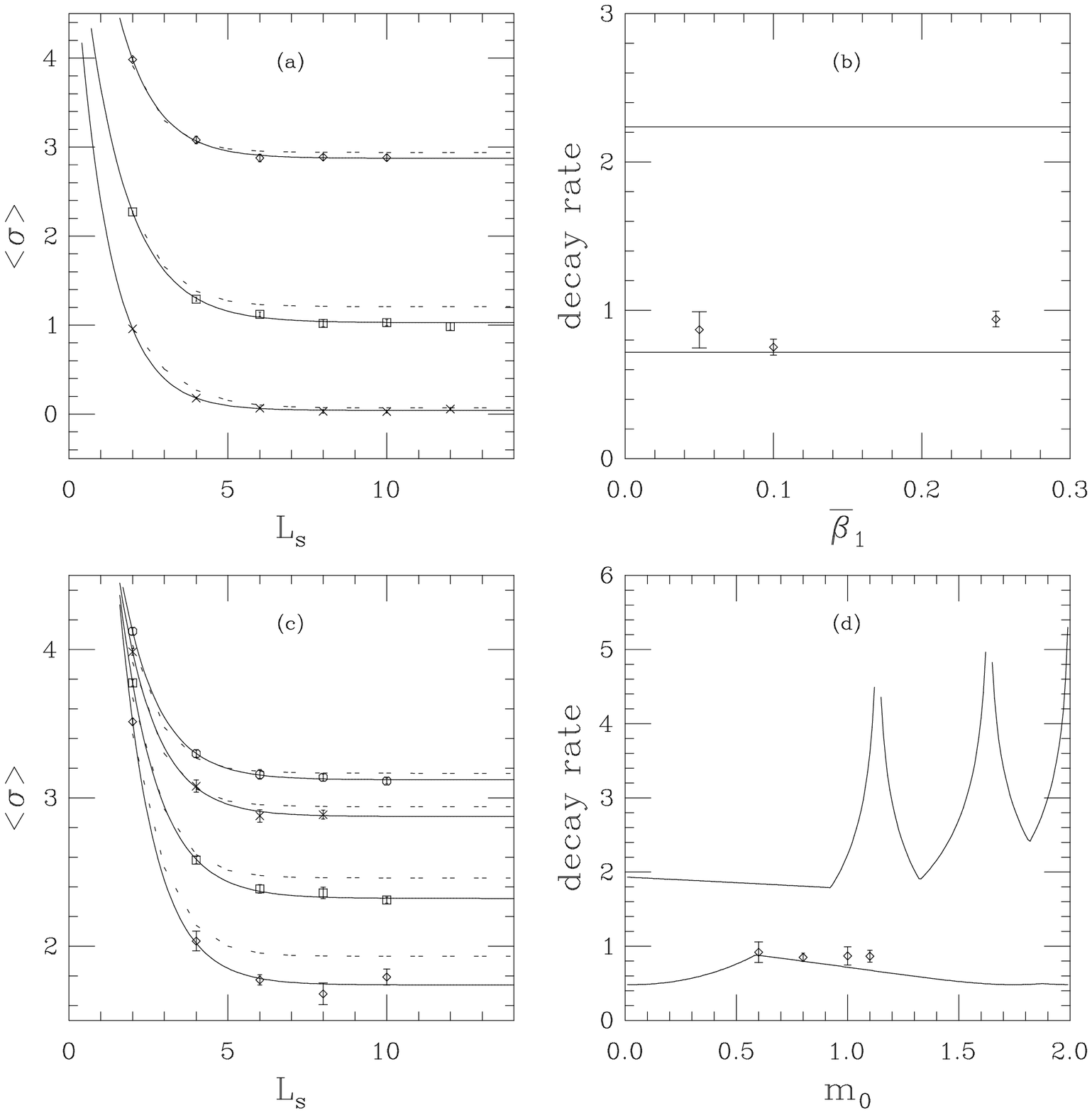}{\seven}{\figsizea} 
\null \vskip -0.8 truein
\figure{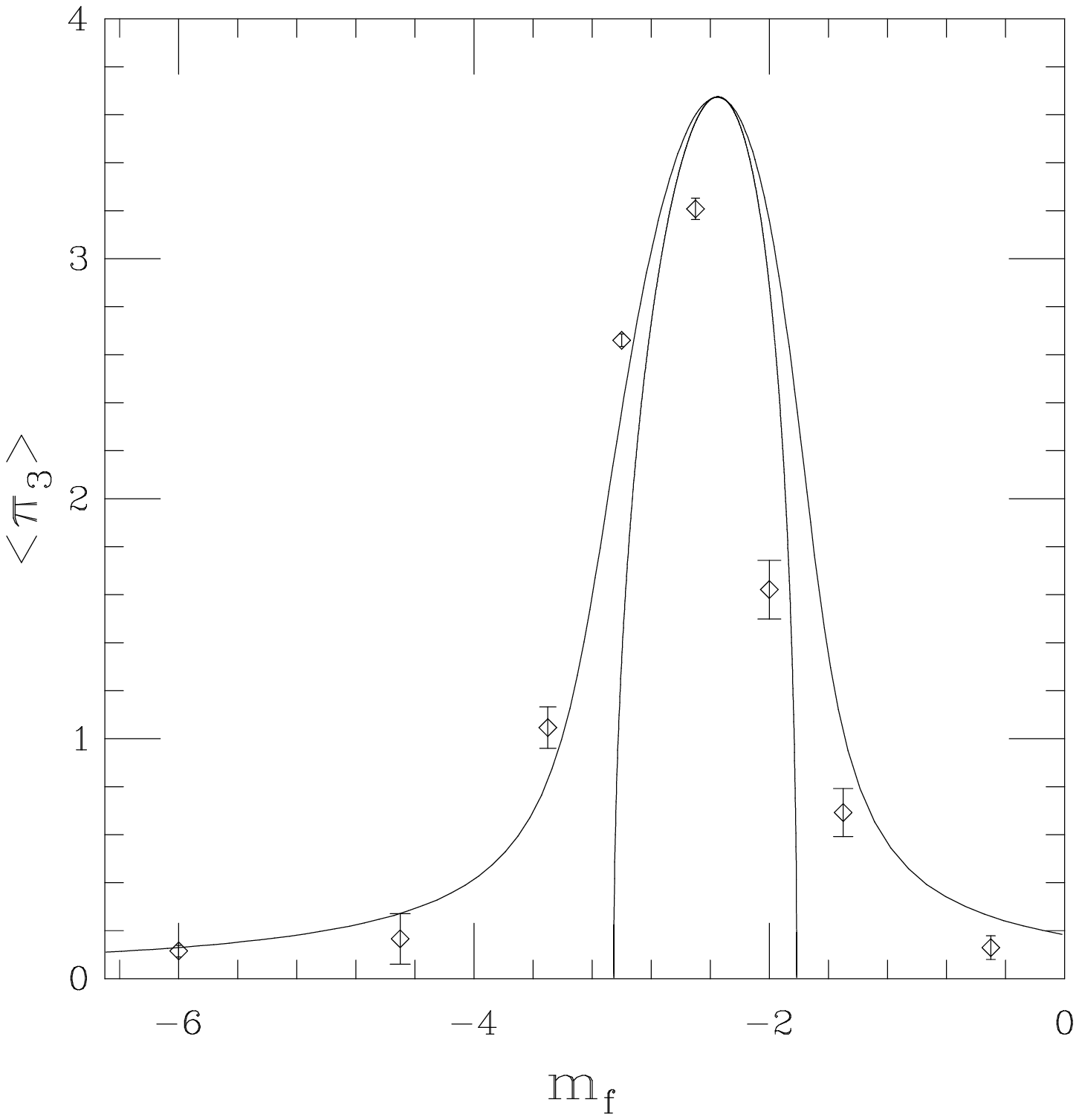}{\eight}{\figsizeb} 
\fi 
%
\if \epsfpreprint N \eject 
\section* {Figure Captions.}
\noindent{\bf Figure 1:} \one 
\noindent{\bf Figure 2:} \two 
\noindent{\bf Figure 3:} \three 
\noindent{\bf Figure 4:} \four 
\noindent{\bf Figure 5:} \five
\noindent{\bf Figure 6:} \six
\noindent{\bf Figure 7:} \seven
\noindent{\bf Figure 8:} \eight
\fi

\begin{thebibliography}{9}

\bibitem{Kaplan} D.B. Kaplan, Phys. Lett. {\bf B288} (1992) 342;
Nucl. Phys. {\bf B30} (Proc. Suppl.)  (1993) 597.

\bibitem{DWF_reviews} 
R. Narayanan, Nucl. Phys. {\bf B34} (Proc. Suppl.)  (1994) 95; 
M. Creutz, Nucl. Phys. {\bf B42} (Proc. Suppl.)  (1995) 56; 
Y. Shamir, Nucl. Phys. {\bf B47} (Proc. Suppl.)  (1996) 212; 
T. Blum, to appear in the Lattice 98 conference proceedings, hep-lat/9810017.

\bibitem{Furman_Shamir} V. Furman, Y. Shamir, Nucl. Phys. {\bf B439} (1995) 54.

\bibitem{lat98_PMV} 
P. Chen, N. Christ, G. Fleming,
A. Kaehler, C. Malureanu, R. Mawhinney, G. Siegert, C. Sui, P. Vranas,
and Y. Zhestkov, contribution to Lattice 98,
hep-lat/9809159.

\bibitem{lat98_Fleming} 
P. Chen, N. Christ, G. Fleming,
A. Kaehler, C. Malureanu, R. Mawhinney, G. Siegert, C. Sui, P. Vranas,
and Y. Zhestkov, contribution to Lattice 98,
hep-lat/9811013.

\bibitem{lat98_Kaehler} 
P. Chen, N. Christ, G. Fleming,
A. Kaehler, C. Malureanu, R. Mawhinney, G. Siegert, C. Sui, P. Vranas,
and Y. Zhestkov, contribution to Lattice 98, talk by A. Kaehler.

\bibitem{ICHEP_NHC}
P. Chen, N. Christ, G. Fleming,
A. Kaehler, C. Malureanu, R. Mawhinney, G. Siegert, C. Sui, P. Vranas,
and Y. Zhestkov, contribution to ICHEP 98,
hep-lat/9812011.

\bibitem{dpf99_PMV} 
P. Vranas, contribution to DPF 99, hep-lat/9903024.

\bibitem{Bitar_private} 
K.M. Bitar, personal communication with P. Vranas.

\bibitem{Ichinose}
I. Ichinose and K. Nagao, hep-lat/9905001.

\bibitem{lat98_RDM} 
P. Chen, N. Christ, G. Fleming,
A. Kaehler, C. Malureanu, R. Mawhinney, G. Siegert, C. Sui, P. Vranas,
and Y. Zhestkov, contribution to Lattice 98,
hep-lat/9811026.

\bibitem{PMV} P.M. Vranas, Lattice 96, 
Nucl. Phys. {\bf B53} (Proc. Suppl.) (1997) 278;
Phys. Rev. {\bf D57} (1998) 1415.

\bibitem{NN1} R. Narayanan, H. Neuberger, Phys. Lett. {\bf B302} (1993) 62;
Phys. Rev. Lett. {\bf 71} (1993) 3251;
Nucl. Phys. {\bf B412} (1994) 574;
Nucl. Phys. {\bf B443} (1995) 305.

\bibitem{Neuberger_fermions} 
H. Neuberger, Phys. Rev. {\bf D57} (1998) 5417; 
Phys. Lett. {\bf B417} 141 (1998); Phys. Rev. Lett. {\bf 81} (1998) 4060;
hep-lat/9901003; 
U.M. Heller, R. Edwards and R. Narayanan hep-lat/9807017; hep-lat/9811030;
C. Lieu, hep-lat/9811008.

\bibitem{Kikukawa}
Y. Kikukawa, T. Noguchi, hep-lat/9902022.

\bibitem{chiral_scalar}
M. Golterman, K. Jansen, D. Petcher and J. Vink,
Phys. Rev. {\bf D49} (1994) 1606.

\bibitem{Bitar_Vranas} K.M. Bitar and P.M. Vranas, 
Phys.Lett. {\bf B327} (1994) 101;
Nucl. Phys. {\bf B34} (Proc. Suppl.) (1994) 661;
Phys. Rev. {\bf D50} (1994) 3406.

\bibitem{Kogut_NJL1}
S. Hands, A. Kocic, J.B. Kogut, Annals Phys. {\bf 224} (1993) 29;
Nucl. Phys. {\bf B390} (1993) 355.

\bibitem{Kogut_NJL2}
A. Kocic, J.B. Kogut, Phys. Rev. Lett. {\bf 74} (1995) 3109;
Nucl. Phys. {\bf B455} (1995) 229.

\bibitem{Kogut_NJL3}
S. Kim, J.B. Kogut, M.P. Lombardo Nucl. Phys. {\bf B53} (Proc.Suppl)
(1997) 709.

\bibitem{Stag_NJL}A. Ali Khan, M. Gockeler, R. Horsley, P.E.L. Rakow, 
G. Schierholz, H. Stuben,
Nucl. Phys. {\bf B34} (Proc.Suppl.) (1994) 655;
Int. J. Mod. Phys. {\bf C5} (1994) 351.

\bibitem{Aoki} S. Aoki, Phys. Rev. {\bf D30} (1984) 2653.

\bibitem{Aoki_NJL} S. Aoki, S. Boettcher and A. Gocksch,
Phys. Lett. {\bf B331} (1994) 157.

\bibitem{Bitar_Vranas_lat94} K.M. Bitar and P.M. Vranas,
Nucl. Phys. {\bf B42} (Proc. Suppl.)  (1995) 746.

\bibitem{Narayanan_private}
R. Narayanan, personal communication with P. Vranas.

\bibitem{Columbia_private}
RIKEN-BNL-Columbia lattice group, in progress.

\bibitem{Sharpe}
S. Sharpe and R. Singleton, Phys. Rev. {\bf D58} (1998) 074501.

\bibitem{Aoki_Gocksch} 
S. Aoki and A. Gocksch, Phys. Rev. {\bf D45} (1992) 3845.

\bibitem{Bitar_pb} K.M. Bitar, 
Nucl. Phys. {\bf B63} Proc. Suppl. (1998) 829; 
Phys. Rev. {\bf D56} (1997) 2736.

\end{thebibliography}
\end{document}